\documentclass[10pt,english]{article}
\usepackage[T1]{fontenc}
\usepackage[latin1]{inputenc}
\usepackage{graphicx}

\makeatletter

\usepackage{helvet}

\usepackage{geometry}



\usepackage{babel}
\makeatother
\begin{document}

\title{Spin-$\frac{1}{2}$ Heisenberg ladder: variation of entanglement
and fidelity measures close to quantum critical points}

\author{Amit Tribedi and Indrani Bose }

\maketitle
\begin{center}Department of Physics \end{center}

\begin{center}Bose Institute \end{center}

\begin{center}93/1, Acharya Prafulla Chandra Road \end{center}

\begin{center}Kolkata - 700 009, India \end{center}

\begin{abstract}
We consider a two-chain, spin-$\frac{1}{2}$ antiferromagnetic Heisenberg
spin ladder in an external magnetic field $H$. The spin ladder is
known to undergo second order quantum phase transitions (QPTs) at
two critical values, $H_{c1}$ and $H_{c2}$, of the magnetic field.
There are now known examples of strongly coupled (rung exchange interaction
much stronger than nearest-neighbour intrachain exchange interaction)
organic ladder compounds in which QPTs have been experimentally observed.
In this paper, we investigate whether well-known bipartite entanglement
meaures like one-site von Neumann entropy, two-site von Neumann entropy
and concurrence develop special features close to the quantum critical
points. As suggested by an earlier theorem, the first derivatives
of the measures with respect to magnetic field are expected to diverge
as $H\rightarrow H_{c1}$ and $H\rightarrow H_{c2}$. Based on numerical
diagonalization data and a mapping of the strongly-coupled ladder
Hamiltonian onto the $XXZ$ chain Hamiltonian, for which several analytical
results are known, we find that the derivatives of the entanglement
measures diverge as $H\rightarrow H_{c2}$ but remain finite as $H\rightarrow H_{c1}$.
The reason for this discrepancy is analysed. We further calculate
two recently proposed quantum information theoretic measures, the
reduced fidelity and reduced fidelity susceptibility, and show that
these measures provide appropriate signatures of the QPTs occuring
at the critical points $H=H_{c1}$ and $H=H_{c2}$.\\
\\
PACS number(s): 03.67.Mn
\end{abstract}

\section*{I. INTRODUCTION}

Antiferromagnetic (AFM) Heisenberg ladders are examples of interacting
many body systems which exhibit a range of novel phenomena \cite{key-1,key-2}.
An $n$-chain spin ladder consists of $n$ chains coupled by rungs,
the simplest example being a two-chain ladder with $n=2$. The study
of ladders as prototypical many body systems became important after
the discovery of high temperature superconductivity in the strongly
correlated cuprate materials. The dominant electronic and magnetic
properties of the cuprates are associated with the $CuO_{2}$ planes
which have the structure of a square lattice \cite{key-3}. The level
of rigour that can be achieved in the treatment of strong correlation
is less in two dimensions (2d) than in 1d. Ladder models, with structure
interpolating between 1d and 2d, serve as ideal candidates to address
issues related to strong correlation and also to investigate how electronic
and magnetic properties change as one progresses from the chain to
the plane. In undoped ladder models, each site of the ladder is occupied
by a spin (usually of magnitude $\frac{1}{2}$) and the spins interact
via the AFM Heisenberg exchange interaction. In doped ladder models,
some of the spins are replaced by positively charged holes which are
mobile. The Hamiltonian describing the doped systems are the t-J and
Hubbard ladder models \cite{key-1,key-2,key-3}. With the discovery
of a large number of materials having a ladder-like structure, the
study of ladders has acquired considerable importance. The materials
exhibit a range of phenomena including superconductivity in hole-doped
systems, the `odd-even' effect in which the excitation spectrum of
an $n$-chain ladder is gapped (gapless) if n is even (odd) and quantum
phase transitions (QPTs) tuned by an external magnetic field \cite{key-1,key-2,key-3}.
Many of the experimental observations were motivated by theoretical
predictions, superconductivity being a prime example \cite{key-4,key-5,key-6}.

In this paper, we focus on QPTs in a two-leg AFM Heisenberg ladder
(Fig. 1) in an external magnetic field. The Hamiltonian describing
the model is given by \begin{equation}
\mathcal{H}=\sum_{j=1}^{L}[J_{||}(\mathbf{S}_{1,j}.\mathbf{S}_{1,j+1}+\mathbf{S}_{2,j}.\mathbf{S}_{2,j+1})+J_{\bot}\mathbf{S}_{1,j}.\mathbf{S}_{2,j}]-H\sum_{j=1}^{L}(S_{1,j}^{z}+S_{2,j}^{z})\label{1}\end{equation}
where the indices $1$ and $2$ distinguish the lower and upper legs
of the ladder and $j$ labels the rungs. The spins have magnitude
$\frac{1}{2}$ ($|\overrightarrow{\mathbf{S}_{j}}|=\frac{1}{2}$)
and interact via the AFM Heisenberg exchange interaction. The intrachain
and rung exchange couplings are of strengths $J_{||}$ and $J_{\bot}$respectively.
The total number of rungs is $L$ and periodic boundary conditions
are assumed. The factor $g\,\mu_{B}$ ($g$ is the Land\'{e} splitting
factor and $\mu_{B}$ the Bohr magneton) is absorbed in $H$. If $J_{\bot}=0$,
the ladder decouples into two non-interacting spin-$\frac{1}{2}$
Heisenberg chains with no gap to spin excitations. For any arbitrary
$J_{\bot}$$\neq0$, the excitation spectrum acquires a gap (spin
gap). In the strong coupling limit, $J_{\bot}>>J_{||}$, a simple
physical picture of the ground state and the origin of the spin gap
can be given. The spins along the rungs predominantly form singlets
in the ground state. A spin excitation is created by replacing a singlet
by a triplet which propagates along the ladder due to the intrachain
exchange interaction. In first order perturbation theory, the spin
gap $\Delta$ is given by $\Delta\approx J_{\bot}-J_{||}$ separating
the lowest excited state from the dimerized ground state. 

There are now several known strong coupling ladder compounds \cite{key-7}.
Of these, the organic ladder compounds $Cu_{2}(C_{5}H_{12}N_{2})_{2}Cl_{4}$
\cite{key-8}, $(C_{5}H_{12}N)_{2}CuBr_{4}$ \cite{key-9} and $(5IAP)2CuBr_{4}.2H_{2}O$
\cite{key-10} are of special interest because of the experimental
observation of the QPTs in these systems by the tuning of an external
magnetic field. A QPT occurs at $T=0$ and brings about a qualitative
change in the ground state of an interacting many body system at a
specific value $g_{c}$ of the tuning parameter $g$ \cite{key-11}.
QPTs are driven by quantum fluctuations and in the case of second
order transitions, the quantum critical point is associated with scale
invariance and a diverging correlation length. The ground state energy
becomes non-analytic at the critical value $g_{c}$ of the tuning
parameter. If one of the phases is gapped, the gap goes to zero in
a power-law fashion as $g\rightarrow g_{c}$. In the case of a spin
ladder, the external magnetic field $H$ plays the role of the tuning
parameter $g$. There are two critical points, $H_{c1}$ and $H_{c2}$
\cite{key-7,key-8,key-9,key-10}. At $T=0$ and for $0<H<H_{c1}$,
the ladder is in the spin gap phase. In the presence of the magnetic
field, there is a Zeeman splitting of the triplet ($S=1$) excitation
spectrum with the $S^{z}=1$ component having the lowest energy. The
spin gap is now $\Delta-H$. At $H=H_{c1}=\Delta$, the gap closes
and a QPT occurs to the Luttinger liquid (LL) phase characterized
by a gapless excitation spectrum. At the upper critical field $H=H_{c2}$,
there is another QPT to the fully polarized ferromagnetic (FM) state.
The magnetization data exhibit universal scaling behaviour in the
vicinity of $H_{c1}$ and $H_{c2}$, consistent with theoretical predictions
\cite{key-7,key-8,key-9,key-10}. In the gapless regime $H_{c1}<H<H_{c2}$,
the ladder Hamiltonian can be mapped onto an XXZ chain Hamiltonian
the thermodynamic propeties of which can be calculated exactly using
the Bethe Ansatz (BA) \cite{key-7,key-8,key-9,key-10}. The theoretically
computed magnetization versus magnetic field curve is in excellent
agreement with the experimental data. QPTs can be observed in the
organic ladder compounds as the magnitudes of the critical fields
are experimentally accessible.

In recent years, QPTs have been extensively studied in spin systems
using well-known quantum information theoretic measures. A number
of entanglement measures have been identified which develop special
features close to the quantum critical point \cite{key-13,key-14,key-15,key-16,key-17,key-18,key-19}.
It has been shown \cite{key-16} that, in general, a first order QPT
linked to a discontinuity in the first derivative of the ground state
energy, is signalled by a discontinuity in a bipartite entanglement
measure whereas a discontinuity or a divergence in the first derivative
of the entanglement measure marks a second-order phase transition
characterized by a discontinuity/divergence in the second derivative
of the ground state energy. Another measure which provides a signature
of QPTs is that of ground state fidelity \cite{key-20,key-21}. The
utility of the measure and a related measure, fidelity susceptibility,
has been explored in a number of studies \cite{key-22,key-23,key-24,key-25,key-26,key-27,key-28}.
Fidelity, a concept borrowed from quantum information theory, is defined
as the overlap modulus between ground states corresponding to slightly
different Hamiltonian parameters. The fidelity typically drops in
an abrupt manner at a critical point indicating a dramatic change
in the nature of the ground state wave function. This is accompanied
by a divergence of the fidelity susceptibility. In these approaches,
the fidelity measure involves global ground states. Recently, the
concept reduced fidelity (RF) (also called partial state fidelity)
has been developed, which relates to the fidelity of a subsystem \cite{key-25,key-29,key-30,key-31,key-32,key-33},
along with the associated notion of reduced fidelity susceptibility
(RFS). Using the RF and RFS measures, QPTs have been studied in spin
models like the Lipkin-Meshkov-Glick model \cite{key-30,key-31},
the transverse field Ising model in 1d \cite{key-32} and the spin-$\frac{1}{2}$
dimerized Heisenberg chains \cite{key-33}. In this paper, we use
some well-known bipartite entanglement measures, which include one-site
entanglement, two-site entanglement and concurrence, for the study
of QPTs in the $S=\frac{1}{2}$ two-leg AFM Heisenberg ladder (Fig.
$1$) described by the Hamiltonian given in Eq. $(1)$. We show that
the entanglement measures develop characteristic features close to
the quantum critical point $H=H_{c2}$ but not at the critical point
$H=H_{c1}$. We next show that the measures based on the RF and RFS
signal the occurence of QPTs at both the critical points $H=H_{c1}$
and $H_{c2}$.

\section*{II. ENTANGLEMENT AND FIDELITY MEASURES PROBING QPTS}

We first define the various entanglement and fidelity measures which
provide the basis of our calculations. The single-site von Neumann
entropy, a measure of the entanglement of a single spin with the rest
of the system, is given by

\begin{equation}
S(i)=-Tr\rho(i)\, log_{2}\,\rho(i)\label{2}\end{equation}
where $\rho(i)$ is the single-site reduced density matrix \cite{key-14,key-19}.
The two-site entanglement $S(i,j)$ is a measure of the entanglement
between two separate spins, at sites $i$ and $j$, and the rest of
the spins \cite{key-18,key-19}. Let $\rho(i,j)$ be the reduced density
matrix for the two spins, obtained from the full density matrix by
tracing out the spins other than the ones at sites $i$ and $j$.
The two-site entanglement is given by the von Neumann entropy

\begin{equation}
S(i,j)=-Tr\rho(i,j)\, log_{2}\,\rho(i,j)\label{3}\end{equation}
In a translationally invariant system, $S$ depends only on the distance
$n=|j-i|$. A knowledge of the two-site reduced density matrix enables
one to calculate concurrence, a measure of entanglement between two
spins at sites $i$ and $j$ \cite{key-34,key-35}. Let $\rho(i,j)$
be defined as a matrix in the standard basis $\{\left|\uparrow\uparrow\right\rangle ,\:\left|\uparrow\downarrow\right\rangle ,\:\left|\downarrow\uparrow\right\rangle \:\left|\downarrow\downarrow\right\rangle \}$.
One can define the spin-reversed density matrix as $\widetilde{\rho}=(\sigma_{y}\otimes\sigma_{y})\:\rho^{*}\:(\sigma_{y}\otimes\sigma_{y})$,
where $\sigma_{y}$ is the Pauli matrix. The concurrence $C$ is given
by $C=max\{\lambda_{1}-\lambda_{2}-\lambda_{3}-\lambda_{4},\,0\}$
where $\lambda_{i}$'s are square roots of the eigenvalues of the
matrix $\rho\,\widetilde{\rho}$ in descending order. $C=0$ implies
an unentangled state whereas $C=1$ corresponds to maximum entanglement. 

The fidelity $F$ is given by the modulus of the overlap of normalized
ground state wave functions $\left|\psi_{0}(\lambda)\right\rangle $
and $\left|\psi_{0}(\lambda+\delta\lambda)\right\rangle $ for closely
spaced Hamiltonian parameter values $\lambda$ and $\lambda+\delta\lambda$
\cite{key-20,key-21,key-22}.

\begin{equation}
F(\lambda,\:\lambda+\delta\lambda)=|\left\langle \psi_{0}(\lambda)|\psi_{0}(\lambda+\delta\lambda)\right\rangle |\label{4}\end{equation}
Eq. (4) gives a definition of the global fidelity. The reduced fidelity
(RF) \cite{key-29,key-30,key-31,key-32,key-33} refers to a subsystem
and is defined to be the overlap between the reduced density matrices
$\rho\equiv\rho(h)$ and $\widetilde{\rho}\equiv\rho(h+\delta)$ of
the ground states $\left|\phi_{0}(h)\right\rangle $ and $\left|\phi_{0}(h+\delta)\right\rangle $,
$h$ and $h+\delta$ being two closely spaced Hamiltonian parameter
values. The RF is 

\begin{equation}
F_{R}(h,\: h+\delta)=Tr\,\sqrt{\rho^{\frac{1}{2}}\,\widetilde{\rho}\,\rho^{\frac{1}{2}}}\label{5}\end{equation}

We now compute the different entanglement and fidelity measures for
the ladder Hamiltonian given in Eq. $(1)$. The external magnetic
field $H$ serves as the Hamiltonian parameter. One notes that the
z-component, $S_{tot}^{z}=\sum_{j=1}^{L}(S_{1,j}^{z}+S_{2,j}^{z})$,
of the total spin is a conserved quantity. Using this fact, the Hamiltonian
is diagonalized for different values of $L$ with the help of the
numerical diagonalization package TITPACK \cite{key-36}. We take
$J_{\bot}=13\, K\mbox{ and }J_{||}=1.15\, K$ which are the approximate
values of the rung and intrachain exchange couplings in the AFM compound
$(5IAP)2CuBr_{4}.2H_{2}O$ \cite{key-10}. We determine the ground
state as well as the three lowest excited state energies for different
values of $H$ with $L$ ranging from $L=2$ to $L=16$. Using the
data, we examine the variation of the fidelity $F(H,\: H+\delta)$
(Eq. $(4)$) with increasing magnetic field strength $H$ and $\delta=.001$.
We observe sharp drops in $F(H,\: H+\delta)$ at $H_{C1}^{L}=\Delta_{L}$
(inset of Fig. $2$), where $\Delta_{L}$ is the spin gap, i.e., the
difference in the energies of the first excited and ground states.
A polynomial fitting of the $\Delta_{L}$ versus $\frac{1}{L}$ data
points yields $\Delta_{L}\approx11.8416+.9739(\frac{1}{L})+.6621(\frac{1}{L})^{2}$.
In the thermodynamic limit $L\rightarrow\infty$, the critical field
is thus $H_{c1}=\Delta_{\infty}\approx11.8416$. In the case of the
strong coupling ladder ($J_{\bot}>>J_{||}$), the critical field $H_{c1}\simeq J_{\bot}-J_{||}$
to the first order in perturbation theory \cite{key-37}. The fully
polarized FM ground state ($H>H_{c2}$) becomes unstable when the
lowest energy of the spin waves falls below the energy of the polarized
state. The magnitude of $H_{c2}$ can be calculated exactly as $H_{c2}=J_{\bot}+2\, J_{||}$.
The estimates of $H_{c1}$ and $H_{c2}$ are in close agreement with
the experimental results \cite{key-2,key-7,key-37}. The numerical
diagonalization data reproduces the exact value of $H_{c2}$ in the
thermodynamic limit. We further obtain the variation of the magnetization
$m(H)$ and its first derivative $\frac{dm}{dH}$ with $H$ in the
thermodynamic limit adopting the extrapolation procedures outlined
in \cite{key-38}. The magnetization $m(H)$ is the average magnetization
per site and because of translational invariance $m(H)=\left\langle S_{i}^{z}\right\rangle $.
At $T=0$, the expectation value is calculated in the ground state.
The inset of Fig. $(3)$ shows that the derivative $\frac{dm}{dH}$
tends to diverge as $H\rightarrow H_{c1}$ and $H_{c2}$. This is
consistent with the existence of a square root singularity in $m(H)$
in the vicinity of the quantum critical points $H_{c1}$ and $H_{c2}$
\cite{key-7,key-37}. Since second order QPTs occur at the critical
fields $H_{c1}$ and $H_{c2}$, the first derivatives of the entanglement
measures, $S(i)$, $S(i,j)$ and $C$, with respect to the tuning
parameter $H$ may exhibit a discontinuity or a divergence as the
critical points are approached \cite{key-16}. We compute the various
first derivatives to ascertain that this feature of critical point
transitions holds true in the case of the spin ladder.

The single-site reduced density matrix $\rho(i)$ can be written in
terms of the spin expectation value $\left\langle S_{i}^{z}\right\rangle $
as \cite{key-31}

\begin{equation}
\rho(i)=\left(\begin{array}{cc}
\frac{1}{2}+\left\langle S_{i}^{z}\right\rangle  & 0\\
0 & \frac{1}{2}-\left\langle S_{i}^{z}\right\rangle \end{array}\right)\label{6}\end{equation}
in the $\left|\uparrow\right\rangle $ ,$\:\left|\downarrow\right\rangle $
basis. From Eq. $(2)$, 

\begin{equation}
S(i)=-\sum_{i}\lambda_{i}\, log_{2}\,\lambda_{i}\label{7}\end{equation}
where the $\lambda_{i}$ 's are the two diagonal elements of $\rho(i)$.
Fig. $3$ shows the variation of $\frac{dS(i)}{dH}$ with $H$. It
is observed that unlike $\frac{dm}{dH}$ , $\frac{dS(i)}{dH}$ tends
to diverge only near $H_{c2}$, while it approaches a finite value
close to $H_{c1}$. The values of $H_{c1}$ and $H_{c2}$ are $H_{c1}=11.8416\: K$
and $H_{c2}=15.3\: K$ as obtained from numerical diagonalization
data. The strongly coupled ladder model in high magnetic field can
be mapped onto a $1d$ $XXZ$ AFM Heisenberg chain with an effective
Hamiltonian \cite{key-37,key-39}

\begin{equation}
\mathcal{H}_{eff}=J_{||}\sum_{j=1}^{L}[\widetilde{S}_{j}^{x}\widetilde{S}_{j+1}^{x}+\widetilde{S}_{j}^{y}\widetilde{S}_{j+1}^{y}+\frac{1}{2}\widetilde{S}_{j}^{z}\widetilde{S}_{j+1}^{z}-\widetilde{H}\sum_{j=1}^{L}\widetilde{S}_{j}^{z}\label{8}\end{equation}
where $\widetilde{H}=H-J_{\bot}-\frac{J_{||}}{2}$ is an effective
magnetic field and $\widetilde{S}_{j}^{\alpha}$ 's $(\alpha=x,\, y,\, z)$
are pseudo spin-$\frac{1}{2}$ operators which can be expressed in
terms of the original spin operators. There are several exact results
known for the $XXZ$ spin-$\frac{1}{2}$ chain in a magnetic field
\cite{key-40,key-41}. In particular, the zero temperature magnetization
$m(H)$ close to the quantum critical points is given by the expressions
(we use the symbol $H$ instead of $\widetilde{H}$)

\begin{equation}
m(H)\sim\frac{\sqrt{2}}{\pi}\sqrt{(H-H_{c1})/J_{||}},\; H>H_{c1}\label{9}\end{equation}

\begin{equation}
m(H)\sim1-\frac{\sqrt{2}}{\pi}\sqrt{(H_{c2}-H)/J_{||}},\; H<H_{c2}\label{10}\end{equation}
Similar expressions are obtained in the case of an integrable spin
ladder model with the help of the thermodynamic BA \cite{key-7}.
Using the analytic expressions of $m(H)$ in Eqs. $(9)$ and $(10)$,
the first derivative of single-site von Neumann entropy with respect
to magnetic field $H$, $\frac{dS(i)}{dH}$, can be calculated analytically
from Eqs. $(2)$ and $(6)$. Again, the derivative diverges near $H_{c2}$
(Fig. $4$) but not as the quantum critical point $H_{c1}$ is approached,
consistent with numerical results. The values of $H_{c1}$ and $H_{c2}$
are $H_{c1}=J_{\bot}-J_{||}=11.85\: K$ and $H_{c2}=J_{\bot}+2\, J_{||}=15.3\: K$.
The estimate of $H_{c1}$ is from first-order perturbation theory.

The correlation functions of the S=$\frac{1}{2}$ $XXZ$ chain in
a magnetic field are known \cite{key-42} in the gapless phase $H_{c1}<H<H_{c2}$.
In terms of the original spin operators, these are given by

\begin{equation}
\left\langle S_{1}^{z}(r)S_{1}^{z}(0)\right\rangle =\frac{m^{2}}{4}+\frac{1}{r^{2}}+cos(2\pi mr)\left(\frac{1}{r}\right)^{2K}\label{11}\end{equation}

\begin{equation}
\left\langle S_{1}^{+}(r)S_{1}^{-}(0)\right\rangle =cos[\pi(1-2m)r]\left(\frac{1}{r}\right)^{\frac{2K+1}{2K}}+cos(\pi r)\left(\frac{1}{r}\right)^{\frac{1}{2K}}\label{12}\end{equation}
where $K$ is the LL exponent. For simplicity, we have dropped some
prefactors (constants) in the terms appearing in Eqs $(11)$ and $(12)$.
The expressions for the correlation functions are utilized to study
the variation of the two-site entanglement $S(i,j)$ and concurrence
$C_{i,i+1}$ with respect to the magnetic field. These quantities
can be computed from the two-site reduced density matrix $\rho(i,j)$
which, in terms of the spin expectation values and correlation functions,
is given by \cite{key-43}\begin{equation}
\rho(i,j)=\left(\begin{array}{cccc}
\frac{1}{4}+\left\langle S_{i}^{z}\right\rangle +\left\langle S_{i}^{z}S_{j}^{z}\right\rangle  & 0 & 0 & 0\\
0 & \frac{1}{4}-\left\langle S_{i}^{z}S_{j}^{z}\right\rangle  & \left\langle S_{i}^{x}S_{j}^{x}\right\rangle +\left\langle S_{i}^{y}S_{j}^{y}\right\rangle  & 0\\
0 & \left\langle S_{i}^{x}S_{j}^{x}\right\rangle +\left\langle S_{i}^{y}S_{j}^{y}\right\rangle  & \frac{1}{4}-\left\langle S_{i}^{z}S_{j}^{z}\right\rangle  & 0\\
0 & 0 & 0 & \frac{1}{4}-\left\langle S_{i}^{z}\right\rangle +\left\langle S_{i}^{z}S_{j}^{z}\right\rangle \end{array}\right)\label{13}\end{equation}
$S(i,j)$ is given by

\begin{equation}
S(i,j)=-\sum_{i}\epsilon_{i}\, log_{2}\,\epsilon_{i}\label{14}\end{equation}
where $\epsilon_{i}$'s are the eigenvalues of $\rho(i,j)$. Using
equation $(9)$, $(10)$, $(11)$ $(12)$ and $(13)$, the first derivative
of $S(i,j)$ with respect to $H$ is calculated near both the critical
points (Fig. $5$). The derivative diverges near $H_{c2}$ but approaches
a finite value close to $H_{c1}$. The n.n. concurrence can be written
as \cite{key-14,key-34,key-35}\begin{equation}
C_{i,i+1}=2\: Max[0,\:\left|\rho_{23}(i,i+1)\right|-\sqrt{\rho_{11}(i,i+1)\rho_{44}(i,i+1)}]\label{15}\end{equation}
Fig. $6$ shows the derivative of $C_{i,i+1}$ with respect to $H$
versus $H$. The derivative, as in the case of one-site and two-site
entanglement measures, diverges as $H\rightarrow H_{c2}$ but has
a finite value as $H\rightarrow H_{c1}$ .

Lastly, we probe the existence of special features, if any, near the
QCPs exhibited by the one-site RF \cite{key-25,key-29,key-30,key-31,key-32,key-33}
defined in Eq. $(5)$. The reduced fidelity susceptibility (RFS) is
defined to be\begin{equation}
\chi_{R}(H)=lim_{\delta\rightarrow0}\frac{-2\: ln\:\mathcal{F}_{R}(H,\: H+\delta)}{\delta^{2}}\label{16}\end{equation}
Figs. $(7)$ and $(8)$ show that the RF $\mathcal{F}_{R}(H,\: H+\delta)$
drops sharply at the quantum critical points ($insets$) and the associated
RFS, $\chi_{R}(H)$, blows up as both the quantum critical points
are approached. This result is in contrast with what is observed in
the case of entanglement measures, where a special feature develops
only in the vicinity of the critical point $H_{c2}$. The calculations
of the RF and the RFS are possible because they involve only local
measures. A calculation of the global fidelity would not have been
possible lacking a knowledge of the true many body ground state.

\begin{figure}
\includegraphics{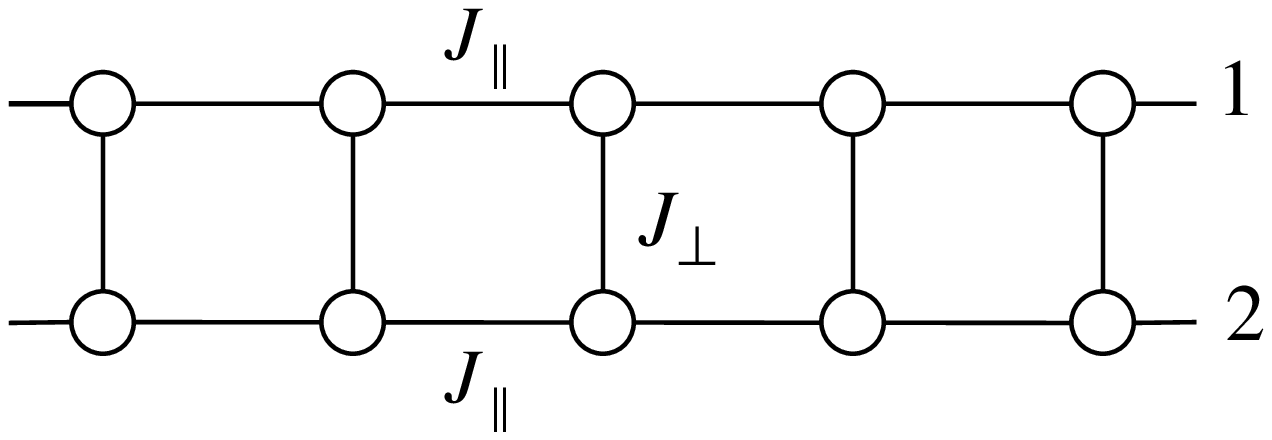}

\textbf{FIG. 1:} A two-chain ladder with rung and intra-chain nearest-neighbour
exchange couplings of strengths $J_{\bot}$ and $J_{||}$ respectively.
The indices $\mathbf{1}$ and $\mathbf{2}$ label the two chains of
the ladder.
\end{figure}
\begin{figure}
\includegraphics{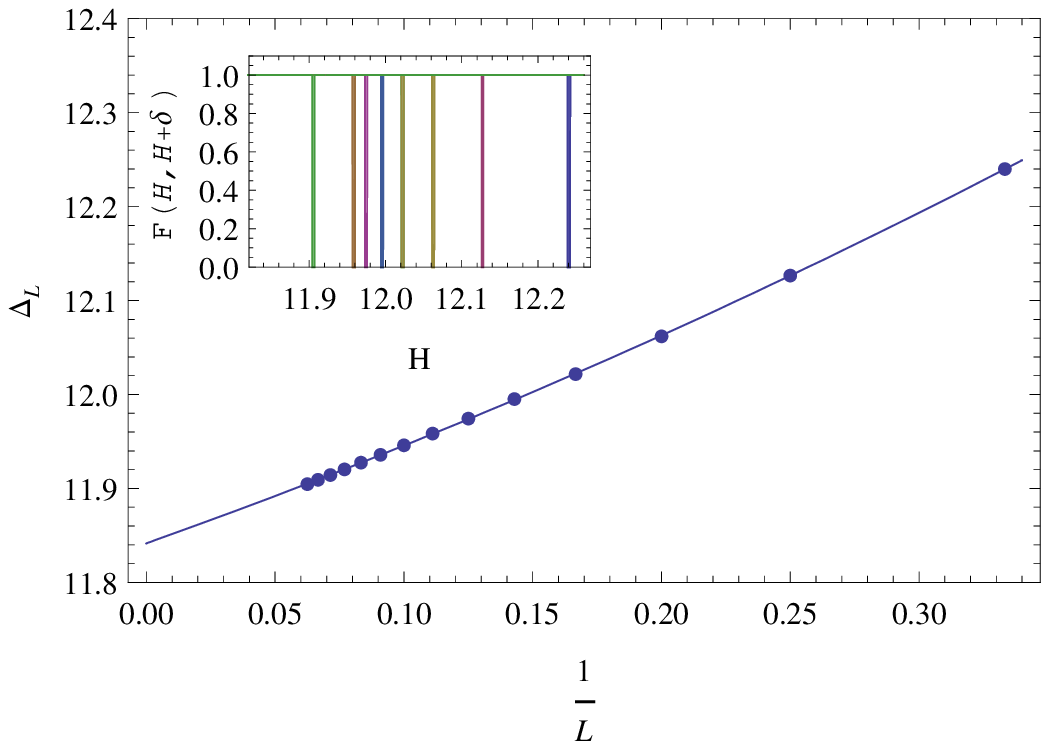}

\textbf{FIG. 2}: Plot of $\Delta_{L}$, the spin gap, versus $\frac{1}{L}$
from numerical diagonalization data of the ladder Hamiltonian (Eq.
$(1)$), $L$ being the number of rungs in the ladder. ($inset$)
ground state fidelity versus magnetic field $H$ for $L=3,4,...16$.
\end{figure}
\begin{figure}
\includegraphics{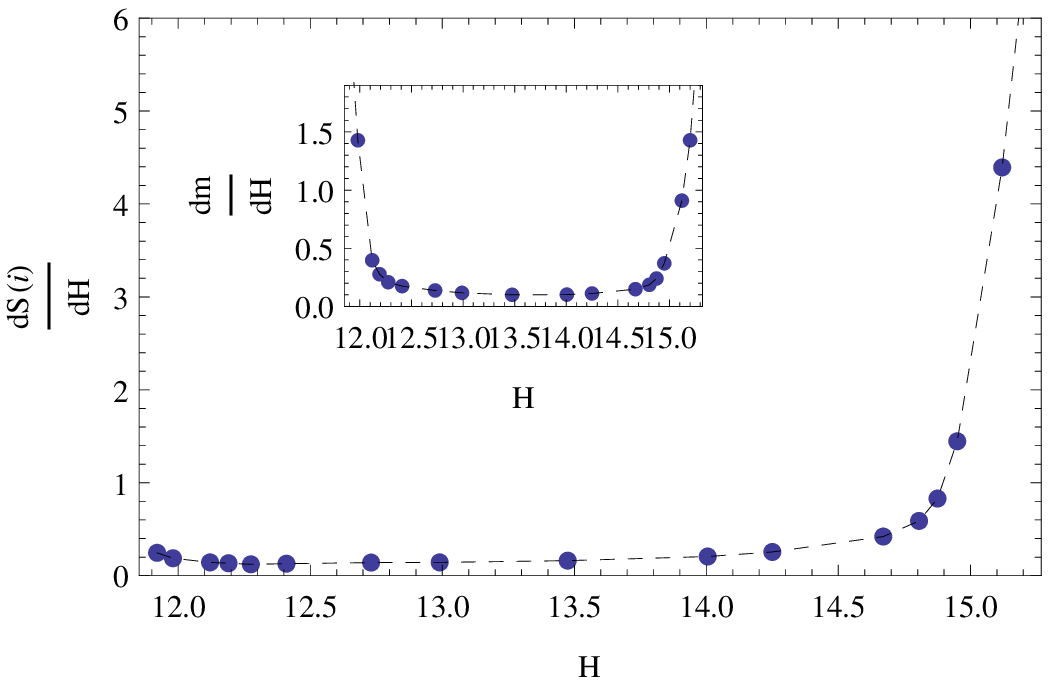}

\textbf{FIG. 3}: Plot of $\frac{dS(i)}{dH}$ versus $H$ (using numerical
diagonalization data); ($inset$) variation of $\frac{dm}{dH}$ with
$H$.
\end{figure}
\begin{figure}
\includegraphics{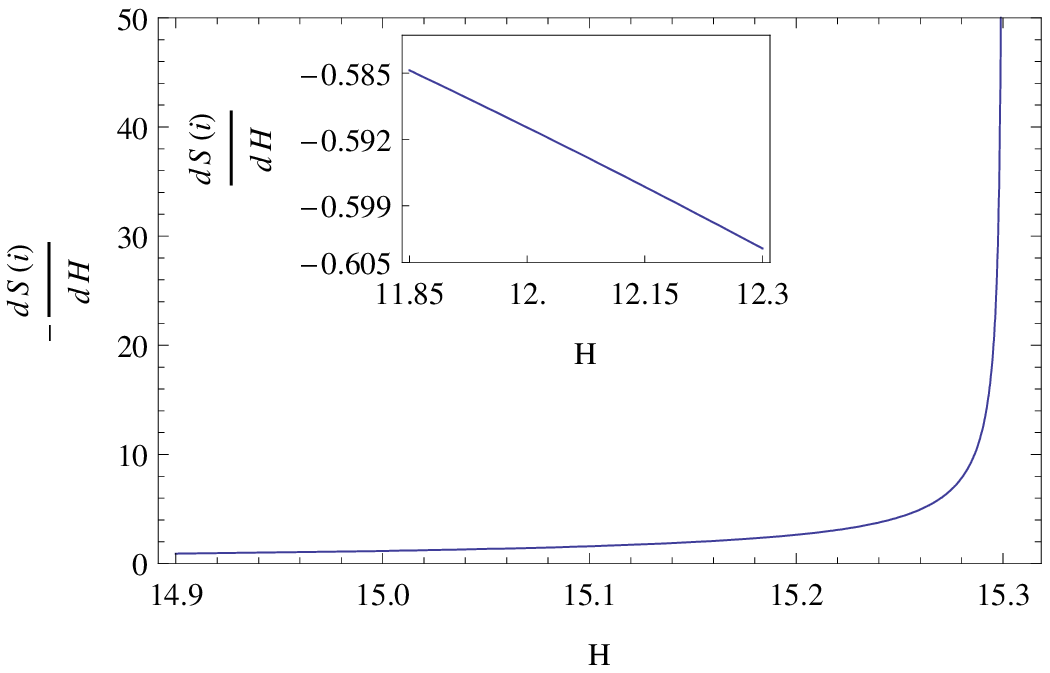}

\textbf{FIG. 4}: Plot of $\frac{dS(i)}{dH}$ versus $H$ near $H=H_{c2}$;
($inset$) plot of $\frac{dS(i)}{dH}$ versus $H$ near $H_{c1}$.
\end{figure}
\begin{figure}
\includegraphics{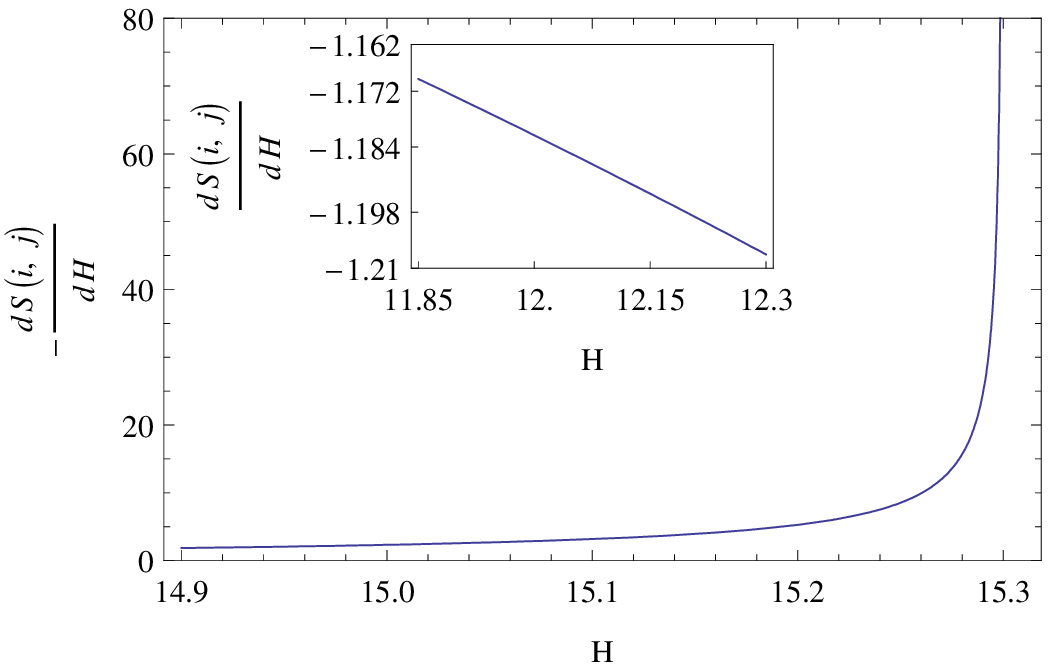}

\textbf{FIG. 5}: Plot of $\frac{dS(i,j)}{dH}$ versus $H$ near $H=H_{c2}$;
($inset$) plot of $\frac{dS(i,j)}{dH}$ versus $H$ near $H_{c1}$.
\end{figure}
\begin{figure}
\includegraphics{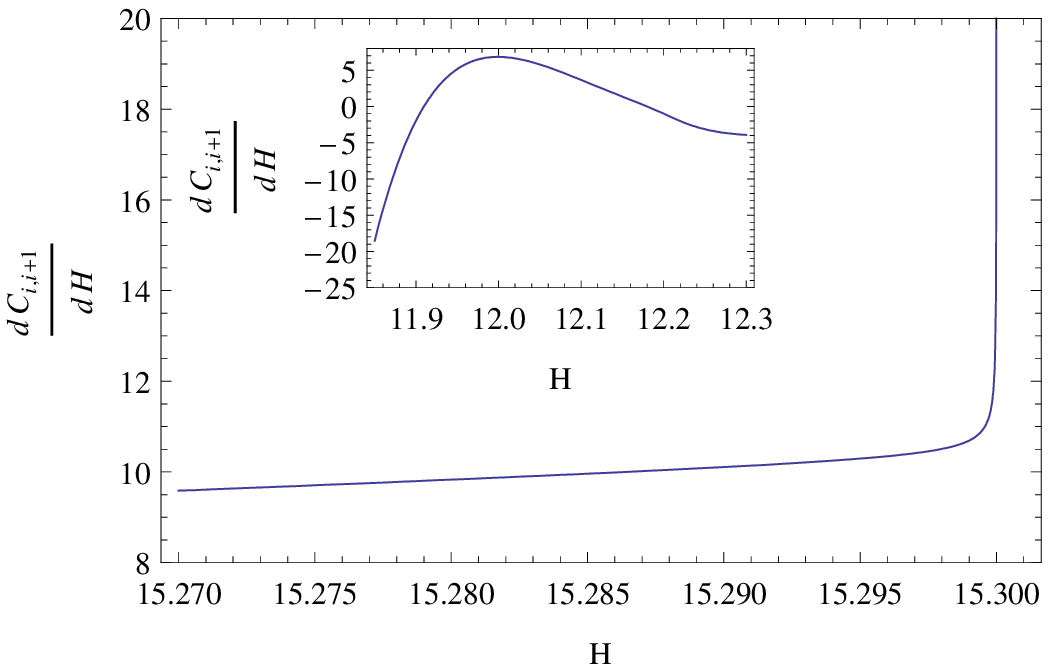}

\textbf{FIG. 6}: Plot of $\frac{dC_{i,i+1}}{dH}$ versus $H$ near
$H=H_{c2}$; ($inset$) plot of $\frac{dC_{i,i+1}}{dH}$ versus $H$
near $H_{c1}$.
\end{figure}
\begin{figure}
\includegraphics{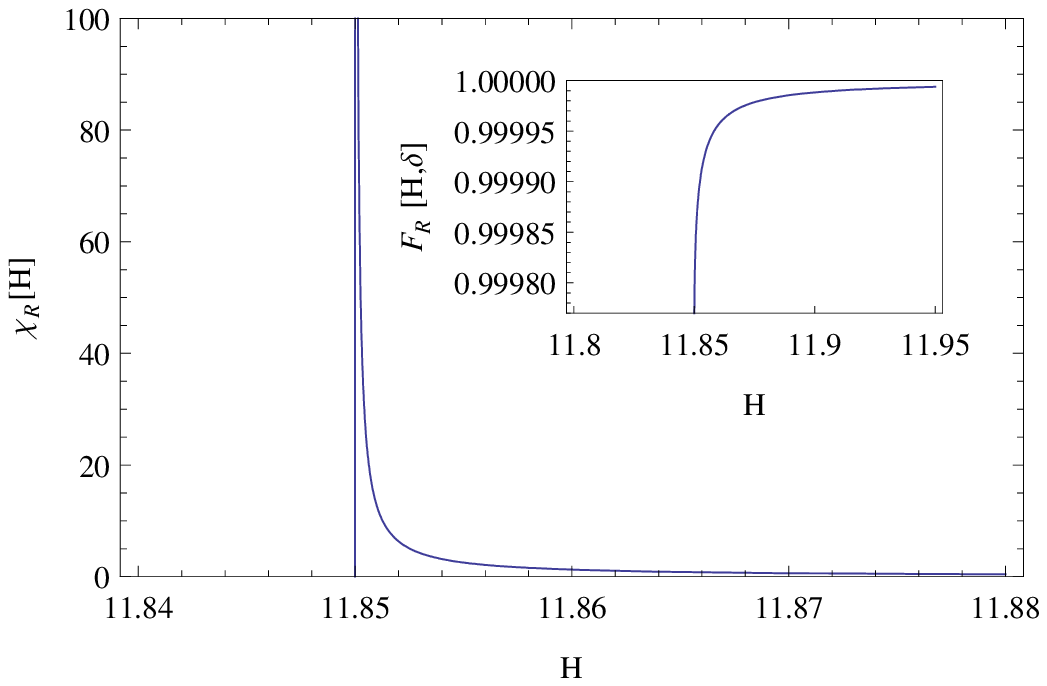}

\textbf{FIG. 7}: Plot of RFS $\chi_{R}(H)$ versus $H$ near $H=H_{c1}$;
($inset$) plot of RF $F_{R}(H,\: H+\delta)$ versus $H$ near $H=H_{c1}$.
\end{figure}
\begin{figure}
\includegraphics{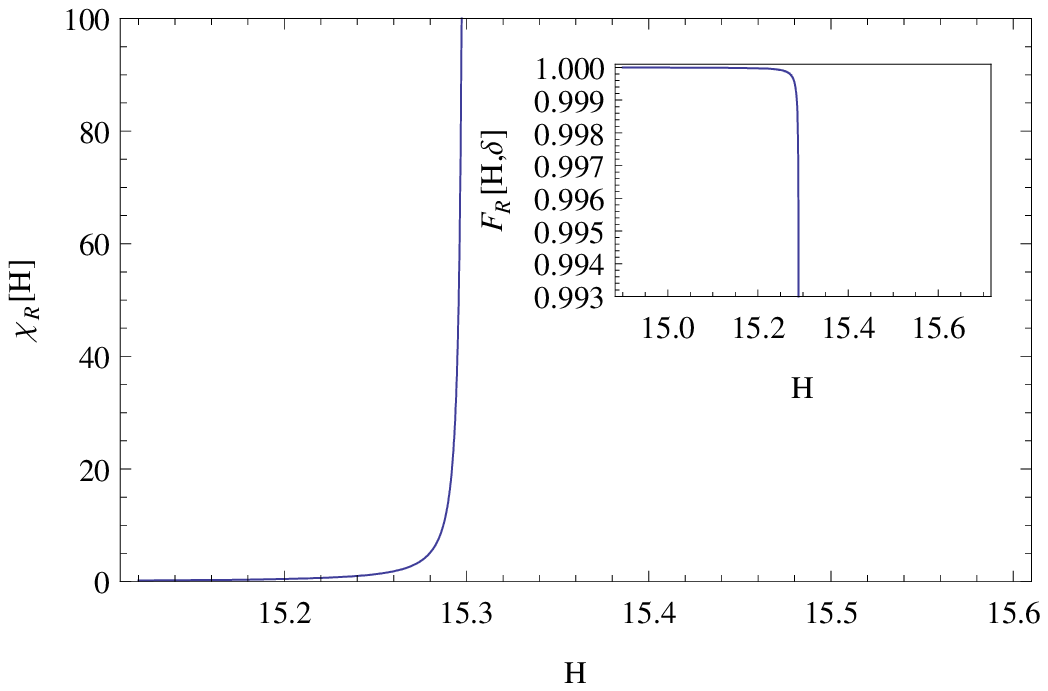}

\textbf{FIG. 8}: Plot of RFS $\chi_{R}(H)$ versus $H$ near $H=H_{c2}$;
($inset$) plot of RF $F_{R}(H,\: H+\delta)$ versus $H$ near $H=H_{c2}$.
\end{figure}

\section*{IV. DISCUSSIONS}

In this paper, we consider a spin-$\frac{1}{2}$, two-chain AFM ladder
in an external magnetic field. The ladder system is known to exhibit
QPTs at two critical values, $H_{c1}$ and $H_{c2}$, of the magnetic
field. The ladder has a rich quantum phase diagram with a gapless
LL phase separating two gapped phases. Both the spin-disordered state
($0<H<H_{c1}$) and the fully polarized FM state ($H>H_{c2}$) constitute
gapped phases. Using a bosonization technique, it has been shown \cite{key-44}
that the spin gaps vanish at the critical points and the spin-spin
correlation functions become long-ranged. As suggested in \cite{key-16,key-17,key-18},
a second order QPT is characterized by a discontinuous/divergent first
derivative of an entanglement measure with respect to the tuning parameter.
Our computations of the first derivatives of the entanglement measures
$S(i)$, $S(i,j)$ and $C_{i,i+1}$ show that these quantities diverge
only as $H\rightarrow H_{c2}$ but remain finite as the other critical
point $H_{c1}$ is approached. As discussed in \cite{key-16}, the
first derivatives of one or more elements of the reduced density matrix
$\rho(i,j)$ with respect to the tuning parameter are expected to
diverge at the critical points. From Eqs. $(11)$-$(13)$, one can
verify that this is the case as $H\rightarrow H_{c1}$ and $H_{c2}$
with the divergent contributions coming from $\rho_{11}(i,j)$ and
$\rho_{44}(i,j)$. The theorem in \cite{key-16} regarding the discontinuity/divergence
of the first derivative of an entanglement measure at a critical point
links the behaviour to that of the first derivative of one or more
elements of $\rho(i,j)$. This is so provided a set of conditions
is satisfied. We find that one of these conditions (condition (b))
is violated in the case of the two-chain ladder as $H\rightarrow H_{c1}$.
This is easily illustrated for the entanglement measure $C_{i,i+1}$
(Eq. $(15)$). The first derivative $\frac{dC_{i,i+1}}{dH}$ involves
terms containing the factor $m(H)\frac{dm(H)}{dH}$ which leads to
a cancellation of singularities as $H\rightarrow H_{c1}$ (see Eq.
$(9)$). This is contrary to condition (b) in \cite{key-16} so that
the theorem is no longer valid. The cancellation of singularities
does not occur as $H\rightarrow H_{c2}$ (see Eq. $(10)$) so that
$\frac{dC_{i,i+1}}{dH}$ signals the occurence of a QPT. In the case
of the single-site entanglement, $S(i)$, similar arguments show that
the cancellation of the singularity occurs as $H\rightarrow H_{c1}$.
The square root singularities in magnetization (Eqs. $(9)$ and $(10)$)
are generic to other AFM systems with spin gap like the spin-$1$
chain in a magnetic field \cite{key-44,key-45,key-46}. Thus, the
behaviour reported in this paper may be a general feature of a class
of gapped $1d$ AFM systems. As shown in our paper, the measures RF
and RFS yield appropriate signatures as both the critical points $H_{c1}$
and $H_{c2}$ are approached and thus appear to be better indicators
of QPTs in the case of systems which violate one or more conditions
of the theorem in \cite{key-16}.

\section*{ACKNOWLEDGMENT}

A. T. is supported by the Council of Scientific and Industrial Research,
India, under Grant No. 9/15 (306)/ 2004-EMR-I. The authors are grateful
to H. Nishimori for sending the full TITPACK program package used
in the present study.


\begin{thebibliography}{10}
\bibitem{key-1}E. Dagotto and T. M. Rice, Science 271, 618 (1996).
\bibitem{key-2}E. Dagotto, Rep. Prog. Phys. 62, 1525 (1999).
\bibitem{key-3}E. Dagotto, Rev. Mod. Phys. 66, 763 - 840 (1994)
\bibitem{key-4}E. Dagotto, J. Riera and D. Scalapino, Phys. Rev. B 45, 5744 (1992).
\bibitem{key-5}S. Gopalan, T. M. Rice and M. Sigrist, Phys. Rev. B 49, 8901 (1994).
\bibitem{key-6}I. Bose and S. Gayen, Phys. Rev. B 48, 10653 (1993).
\bibitem{key-7}M. T. Batchelor, X. W. Guan, N. Oelkers and Z. Tsuboi, Adv. Phys.
56, 465 (2007).
\bibitem{key-8}G. Chaboussant, P.A. Crowell, L. P. L\'{e}vy, O. Piovesana, A. Madouri
and D. Mailly, Phys. Rev. B, 55 3046 (1997).
\bibitem{key-9}B.C. Watson et al., Phys. Rev. Lett., 86 5168 (2001).
\bibitem{key-10}C.P. Landee, M.M. Turnbull, C. Galeriu, J. Giantsidis and F.M. Woodward,
Phys. Rev. B 63 100402 (2001).
\bibitem{key-11}S. Sachdev, Science 288, 475 (2000)
\bibitem{key-12}S. Sachdev, Quantum Phase Transitions, Cambridge University Press,
Cambridge, 1999.
\bibitem{key-13}A. Osterloh, L. Amico, G. Falci, and R. Fazio, Nature (London) 416,
608 (2002).
\bibitem{key-14}T. J. Osborne and M. A. Nielsen, Phys. Rev. A 66, 032110, (2002).
\bibitem{key-15}G. Vidal, J. I. Latorre, E. Rico, and A. Kitaev, Phys. Rev. Lett.
90, 227902 (2003).
\bibitem{key-16}L.-A.Wu, M. S. Sarandy and D. A. Lidar, Phys. Rev. Lett. 93, 250404
(2004).
\bibitem{key-17}T. R. Oliveira, G. Rigolin, M. C. de Oliveira and E. Miranda, Phys.
Rev. Lett. 97, 170401 (2001).
\bibitem{key-18}H.-D. Chen, J. Phys. A 40, 10215 (2007).
\bibitem{key-19}A. Tribedi and I. Bose, Phys. Rev. A 75, 042304 (2007).
\bibitem{key-20}H. T. Quan, Z. Song, X. F. Liu, P. Zanardi, and C. P. Sun, Phys. Rev.
Lett. 96, 140604 (2006).
\bibitem{key-21}P. Zanardi and N. Paunkovi\'{c}, Phys. Rev. E 74, 031123 (2006).
\bibitem{key-22}M. Cozzini, R. Ionicioiu and P. Zanardi, Phys. Rev. B 76, 104420 (2007).
\bibitem{key-23}M. Cozzini, P. Giorda and P. Zanardi, Phys. Rev. B 75, 014439 (2007).
\bibitem{key-24}P. Buonsante and A. Vezzani, Phys. Rev. Lett. 98, 110601 (2007).
\bibitem{key-25}H.-Q. Zhou, e-print arXiv:0704.2945.
\bibitem{key-26}P. Zanardi, M. Cozzini and P. Giorda, J. Stat. Mech.: Theory Exp.
(2007), L02002.
\bibitem{key-27}P. Zanardi, H. T. Quan, X. Wang and C. P. Sun, Phys. Rev. A 75, 032109
(2007).
\bibitem{key-28}S. Chen, L. Wang, S. J. Gu and Y. Wang, Phys. Rev. E 76, 061108 (2007).
\bibitem{key-29}N. Paunkovi\'{c}, P. D. Sacramento, P. Nogueira, V. R. Vieira and
V. K. Dugaev , Phys. Rev. A 77, 052302 (2008).
\bibitem{key-30}H.-M. Kwok, C.-S. Ho and S.- J. Gu, Phys. Rev. A 78, 062302 (2008).
\bibitem{key-31}J. Ma, L. Xu, H. Xiong and X. Wang, arXiv:0805.4062.
\bibitem{key-32}J. Ma, L. Xu and X. Wang, arXiv:0808.1816.
\bibitem{key-33}H.-N. Xiong, J. Ma, Z. Sun and X. Wang, arXiv:0808.1817.
\bibitem{key-34}K. M. O'Connor and W. K. Wootters, Phys. Rev. A 63, 052302 (2001);
W. K. Wootters, Phys. Rev. Lett. 80, 2245 (1998).
\bibitem{key-35}M. C. Arnesen, S. Bose and V. Vedral, Phys. Rev. Lett 87, 017901 (2001);
D. Gunlycke, V. M. Kendon, V. Vedral and S. Bose, Phys. Rev. A 64,
042302 (2001).
\bibitem{key-36}H. Nishimori, AIP Conf. Proc. 248, 269-278 (1992).
\bibitem{key-37}G. Chaboussant et al., Eur. Phys. J. B 6, 167 (1998).
\bibitem{key-38}T. Sakai and M. Takahashi, Phys. Rev. B 43, 13383 (1991).
\bibitem{key-39}F. Mila, Eur. Phys. J. B 6, 201 (1998).
\bibitem{key-40}C. N. Yang and C. P. Yang, Phys. Rev. 150, 327 (1966).
\bibitem{key-41}F. D. M. Haldane, Phys. Rev. Lett. 47, 1840 (1981).
\bibitem{key-42}T. Giamarchi and A. M. Tsvelik, Phys. Rev. B 59, 11398 (1999).
\bibitem{key-43}U. Glaser, H. B\"{u}ttner and H. Fehske, Phys. Rev. A 68, 032318
(2003).
\bibitem{key-44}R. Chitra and T. Giamarchi, Phys. Rev. B 55, 5816 (1997).
\bibitem{key-45}H. J. Schulz, Phys. Rev. B 22, 5274 (1980).
\bibitem{key-46}I. Affleck, Phys. Rev. B 43, 3215 (1991). 
\end{thebibliography}
\end{document}